\documentclass[12pt,preprint]{aastex}

\begin{document}

\title{NGC 602 Environment, Kinematics and Origins\altaffilmark{1,2}}

\author{L. Nigra\altaffilmark{3}}

\email{nigra@astro.wisc.edu}

\author{J. S. Gallagher, III\altaffilmark{3}}

\email{jsg@astro.wisc.edu}

\author{L. J. Smith\altaffilmark{4,5,6}}

\email{lsmith@stsci.edu}

\author{S. Stanimirovi{\'{c}}\altaffilmark{3}}

\email{sstanimi@astro.wisc.edu}

\author{A. Nota\altaffilmark{5}}

\email{nota@stsci.edu}

\and{}

\author{E. Sabbi\altaffilmark{5}}

\email{sabbi@stsci.edu}

\altaffiltext{1}{Based on observations made with the NASA/ESA Hubble Space Telescope, obtained at the Space Telescope Science Institute, which is operated by the Association of Universities for Research in Astronomy, Inc., under NASA contract NAS 5-26555. These observations are associated with program number GO-10396.}

\altaffiltext{2}{Based on observations obtained at the Anglo-Australian Telescope, Siding Spring, Australia.}

\altaffiltext{3}{Department of Astronomy, University of Wisconsin, 475 North Charter Street, Madison, WI 53706-1582, USA}

\altaffiltext{4}{Space Telescope Science Institute, 3700 San Martin Drive, Baltimore, MD 21218, USA}

\altaffiltext{5}{European Space Agency, Research and Scientific Support Department, Baltimore, MD 21218, USA}

\altaffiltext{6}{Department of Physics and Astronomy, University College London, Gower Street, London WC1E 6BT, UK}

\begin{abstract}
The young star cluster NGC~602 and its associated HII region, N90,
formed in a relatively isolated and diffuse environment in the Wing
of the Small Magellanic Cloud. Its isolation from other regions of
massive star formation and the relatively simple surrounding HI shell
structure allows us to constrain the processes that may have led to
its formation and to study conditions leading to massive star formation.
We use images from \emph{Hubble Space Telescope} and high resolution
echelle spectrographic data from the Anglo-Australian Telescope along
with 21-cm neutral hydrogen (HI) spectrum survey data and the shell
catalogue derived from it to establish a likely evolutionary scenario
leading to the formation of NGC~602. We identify a distinct HI cloud
component that is likely the progenitor cloud of the cluster and HII
region which probably formed in blister fashion from the cloud's periphery.
We also find that the past interaction of HI shells can explain the
current location and radial velocity of the nebula. The surrounding
Interstellar Medium is diffuse and dust-poor as demonstrated by a
low visual optical depth throughout the nebula and an average HI density
of the progenitor cloud estimated at 1~cm$^{-3}$. These conditions
suggest that the NGC~602 star formation event was produced by compression
and turbulence associated with HI shell interactions. It therefore
represents a single star forming event in a low gas density region.
\end{abstract}


\section{Introduction}

The young star cluster NGC~602 (J2000 01:29:30, -73:34:0.0) and its
associated nebula, N90 are located in the Wing of the Small Magellanic
Cloud (SMC). This is the relatively diffuse southeastern region which transitions to
the bridge connecting the SMC and the Large Magellanic Cloud (LMC).
Massive star formation is naturally a topic
of perennial interest and the range of conditions under which stars
form at various epochs and host environments makes it a very diverse
field. The NGC~602 cluster is an example of massive star formation
in a region with diffuse interstellar medium (ISM), having the low metallicity
characteristic of the SMC
and without any apparent direct kinetic trigger such as a recent nearby
supernova. Its isolation alone makes it a great laboratory for the
study of star formation since the complexities introduced by kinetic
and optical interaction with other star-forming regions are eliminated.
This combined with low metallicity and relatively quiescent
environment also makes it a candidate for comparison to theoretical
work on primordial HII regions driven by Population III stars (e.g.,
\citealp{abel2007}).

\begin{figure}
\noindent \begin{centering}
\includegraphics[width=7.5cm]{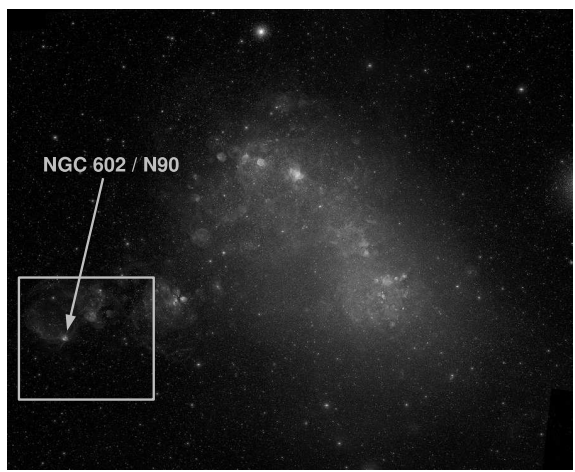} 
\par\end{centering}

\noindent \begin{centering}
\includegraphics[width=7.5cm]{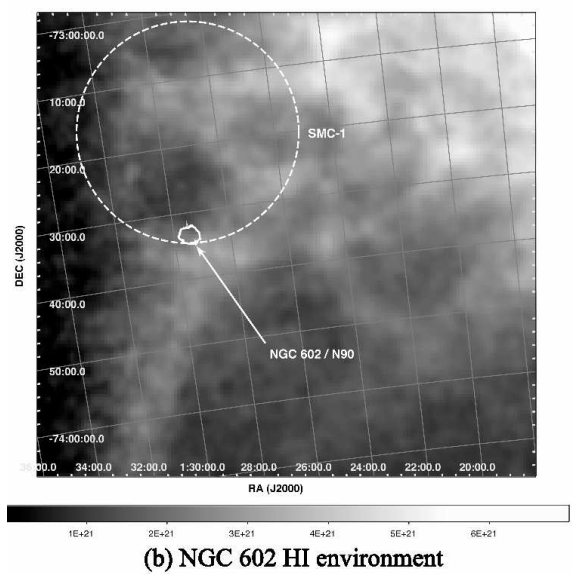}
\par\end{centering}

\caption{(a) NGC~602 and its associated HII region, N90 is located in the
southeastern corner in the Wing of the SMC shown here as an emission
line composite in the upper frame (Composite image credit: F. Winkler/Middlebury
College, the MCELS Team, and NOAO/AURA/NSF). (b) The region around
NGC~602 indicated in the upper frame is expanded and shown in the
lower frame as a HI column density image taken from the data developed
by \citet{stanimirovic1999}. The H$\alpha$ supergiant shell, SMC-1
(faintly visible in the upper frame) is roughly outlined on the HI
image and the nebula, N90 is outlined at SMC-1's southern tip by a
deep H$\alpha$ contour taken from the F658N HST ACS image. The intensity
scale is in units of cm$^{-2}$.\label{fig:NGC-602-and-}}

\end{figure}

NGC~602 consists of a central star cluster surrounded by the N90
HII region. Its location in the eastern periphery of the SMC is indicated
in Figure~\ref{fig:NGC-602-and-} from the Magellanic Cloud Emission
Line Survey (MCELS)
\footnote{SMC image credit: F. Winkler/Middlebury College, the MCELS Team, and NOAO/AURA/NSF. The image was obtained from the web page 
http://www.noao.edu/image\_gallery/html/im0993.html.}.
Also shown in the figure is a HI column density image of the NGC~602
neighborhood taken from the HI survey data used in our study described
in \S~\ref{sec:Data}. The nominal outline of the H$\alpha$ supergiant
shell SMC-1, identified by \citet{meaburn1980}, is shown and at its
southern tip, the nebula N90 is outlined by a H$\alpha$ contour.
We adopt a distance to the SMC of 60~kpc \citep{hilditch2005} and
this results in a transverse scale of 0.29~pc~arcsec$^{-1}$.

\begin{figure}
\begin{centering}
\includegraphics[width=16cm]{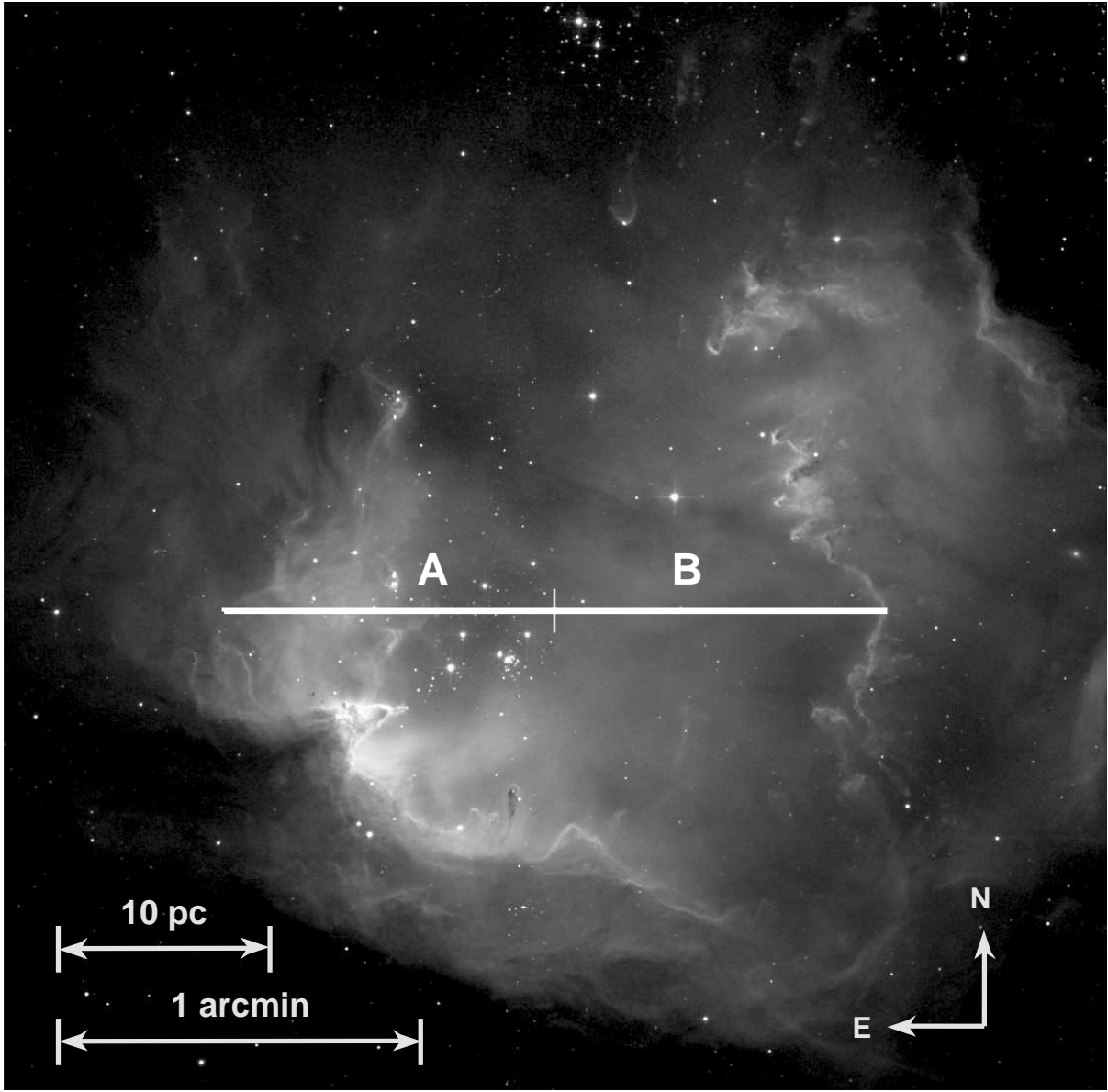} 
\par\end{centering}

\caption{NGC~602/N90 H$\alpha$ image from the HST ACS reveals a rich PDR
structure with many ``elephant trunks'' and prototypical HII region
features. Overlaid on the image are the two slits used with the AAT
UCLES to obtain the HII velocity profile across the nebula shown in
Figure \ref{fig:Slits}. \label{fig:NGC602-Halpha-Slits}}

\end{figure}

The H$\alpha$+{[}NII] \emph{Hubble Space Telescope} (HST) Advanced 
Camera for Surveys (ACS) image of NGC~602 and N90 is shown
in Figure \ref{fig:NGC602-Halpha-Slits} from which the detailed structure
of N90 can be appreciated, but does not do it full justice. A revealing
color composite of all three HST ACS images used in this study has
been published in \citet{carlson2007}. High angular resolution
reveals a very rich structure in the photodissociation regions (PDRs)
along ridges and faces to the east, south and west in NGC~602. The
nebula is host to many prototypical structures found in HII regions
including ``elephant trunks'' and ``pillars of creation''
along the northwestern ridge and what is likely a molecular outcropping
to the southeast. A catalogue and analysis of the many features found
in NGC~602 will be the subject of a new paper.

Also notable are
the many galaxies visible both outside and within the nebula, including
a prominent grand design spiral seen nearly face-on north of the eastern
HII face. The presence of distinct PDRs to the east, south and west,
and the openness toward the north suggests a blister morphology. 
Since the largest dense 
condensation of gas, which also is in proximity to most of the young stars,  
is in the south-southeast part of the nebula, it is most likely that the  
cluster emerged from this region with the first breakout being in the  
projected northern and western directions. 
However, even to the south-southeast, there are galaxies
visible indicating that there is no obvious remaining extended dense
gas concentration there. This transparency suggests a lack of large-scale
dust and molecular concentrations anywhere nearby.

The massive stars associated with NGC~602 are in two groups, one
large group in the southeast and one smaller group to the northwest.
The stellar population of NGC~602 has been studied by \citet{westerlund1964},
\citet{hodge1983}, \citet{hutchings1991} and most recently by \citet{carlson2007}
and \citet{schmalzl2008}. \citeauthor{westerlund1964} estimated
the maximum age at about 10~Myr. \citeauthor{hodge1983} re-analyzed
Westerlund's color-magnitude data and placed the age at 16$\pm5$~Myr.
\citeauthor{hutchings1991} obtained optical and UV spectra of the
OB stars in the cluster and estimated the age at 5~Myr. \citeauthor{carlson2007},
used HST ACS optical and Spitzer IR images and estimated the age of
the OB stars at 4 Myr and suggest that the young objects in the two
ridges are a result of progressive star formation. Progressive star
formation is also suggested by \citet{gouliermis2007}.

The diffuse, quiescent nature of this environment and its low metallicity
raises questions as to how massive stars evidently formed in what appears to be
a relatively isolated region of dense gas. By analyzing optical images, we
look for evidence of the molecular gas substructure left behind in
the cluster's formation. In order to understand the kinematic relationship
of NGC~602 to the larger scale environment, we compare radial velocities of the 
HII gas obtained through high resolution H$\alpha$ longslit spectra to 21-cm HI SMC 
survey data. We also use the kinematics of HI shells in the region to develop 
a simple scenario that could have led to the formation of this cluster.

\section{Data\label{sec:Data}}

High resolution images were taken of NGC~602 with the HST ACS (Program
GO-10248, PI: A. Nota) using the F555W, F814W and F658N filters (V,
I and H$\alpha$+{[}NII], respectively). A description of these images
and their reduction can be found in \citet{carlson2007}. The resolution
achieved in these is 0.05 arcsecond FWHM and this results in a spatial
resolution of about 0.03 pc, or 6300 AU.

High spectral resolution observations of the nebula were obtained
by Linda J. Smith using the University College London Echelle Spectrograph
(UCLES) and the MITLL3 CCD detector at the coud\'e focus of the 3.9~m
Anglo-Australian Telescope (AAT) on 2006 November 05. A slit of dimensions
1 arcsec~$\times$~56 arcsec was used with an interference filter
to isolate a single order containing the H$\alpha$ line. Observations
for two slit positions, spanning the central region of the nebula,
were obtained at a position angle of $90^{\circ}$. The precise slit
positions are shown in Figure \ref{fig:NGC602-Halpha-Slits}. The
total exposure time for each slit position was 2700 sec. The data
were cosmic ray-cleaned, bias-subtracted, and wavelength calibrated.
The spectral resolution was measured to be 5.0~km~sec$^{-1}$ from
ThAr arc lines and each spatial pixel is 0.18~arcsec. 

\begin{figure}
\begin{centering}
\includegraphics[width=16cm]{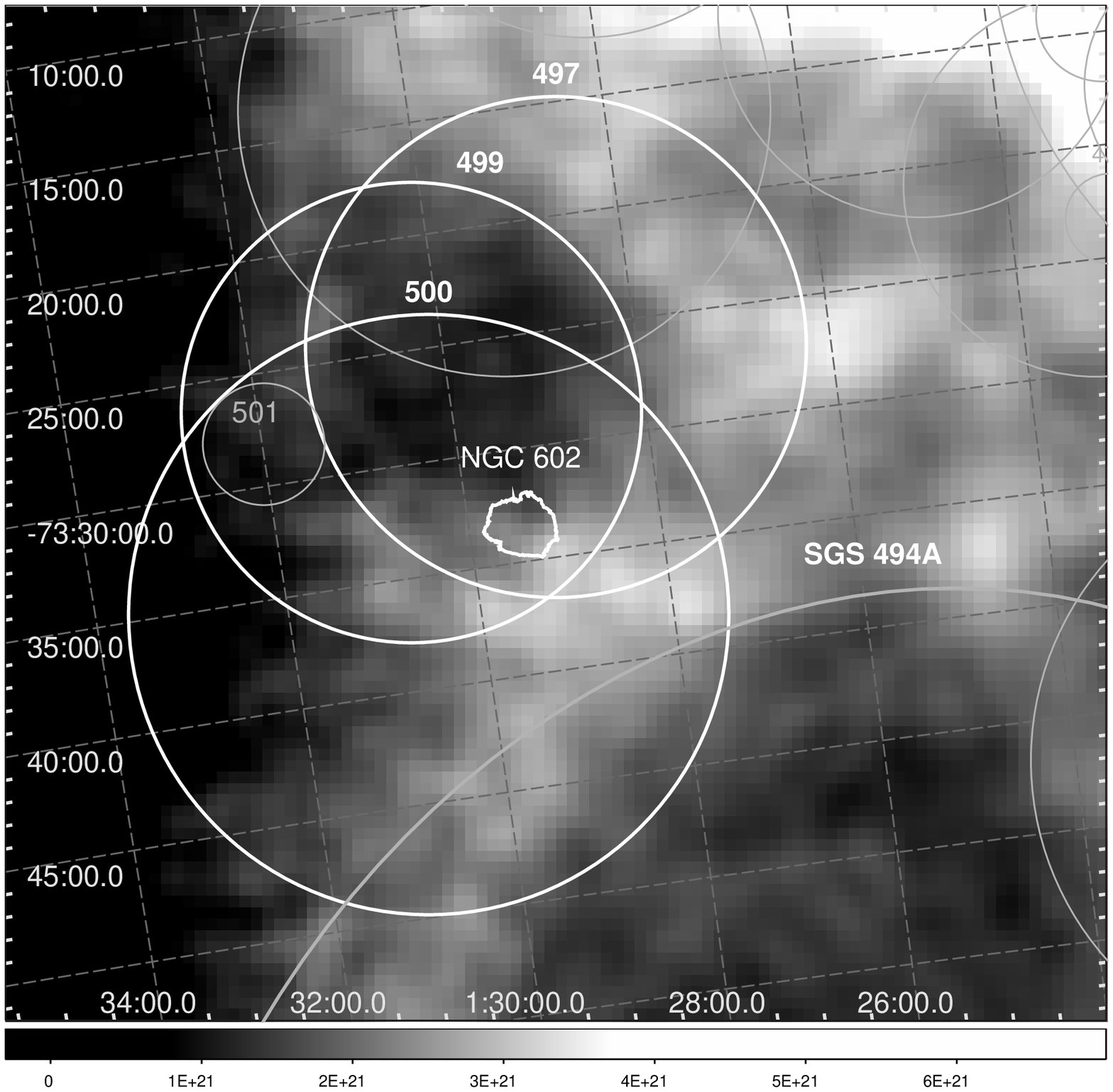} 
\par\end{centering}

\caption{HI total column density image for the vicinity of NGC~602, shown
as an outline of its H$\alpha$ emission region. HI shells identified
by \citet{staveley-smith1997} and \citet{stanimirovic1999} are overlaid
with shells 497, 499 and 500 highlighted. These shells are situated
such that they are candidates for influencing the formation of NGC~602.
Super giant shell SGS~494A may have influenced the present structure.
Coordinates are J2000. The intensity scale is in units of cm$^{-2}$.\label{fig:HI-total-column}}

\end{figure}

The HI structure in the area near NGC~602 was analyzed using a synthesized
21-cm spectral emission map of the SMC described in \citet{stanimirovic1999}.
The data cube consists of $1.6\mbox{\, km\, sec}^{-1}$ heliocentric
velocity slices from 88 to $216\mbox{\, km\, sec}^{-1}$ across the
entire SMC with $30\,\mbox{arcsecond}$ spatial resolution and pixel
size. A complex set of HI shell structures in the SMC has been identified
using these data \citep{staveley-smith1997,stanimirovic1999}. The
integrated HI column density image derived from this map in the area
of the nebula with HI shells overlaid is shown in Figure~\ref{fig:HI-total-column}.

\section{Results}

\subsection{Gas Environment}

\begin{figure}
\begin{centering}
\includegraphics[width=8cm]{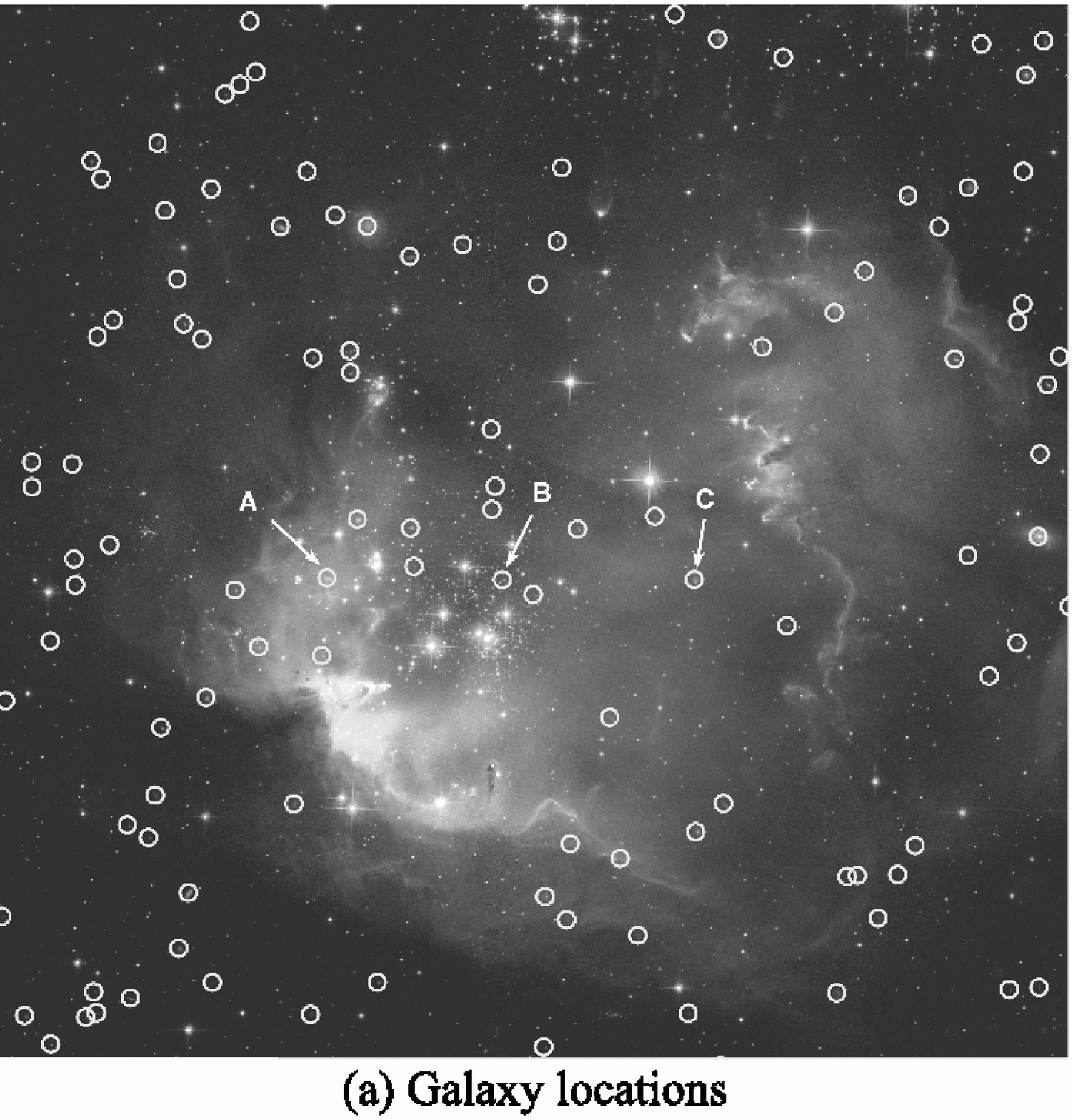}\includegraphics[width=8cm]{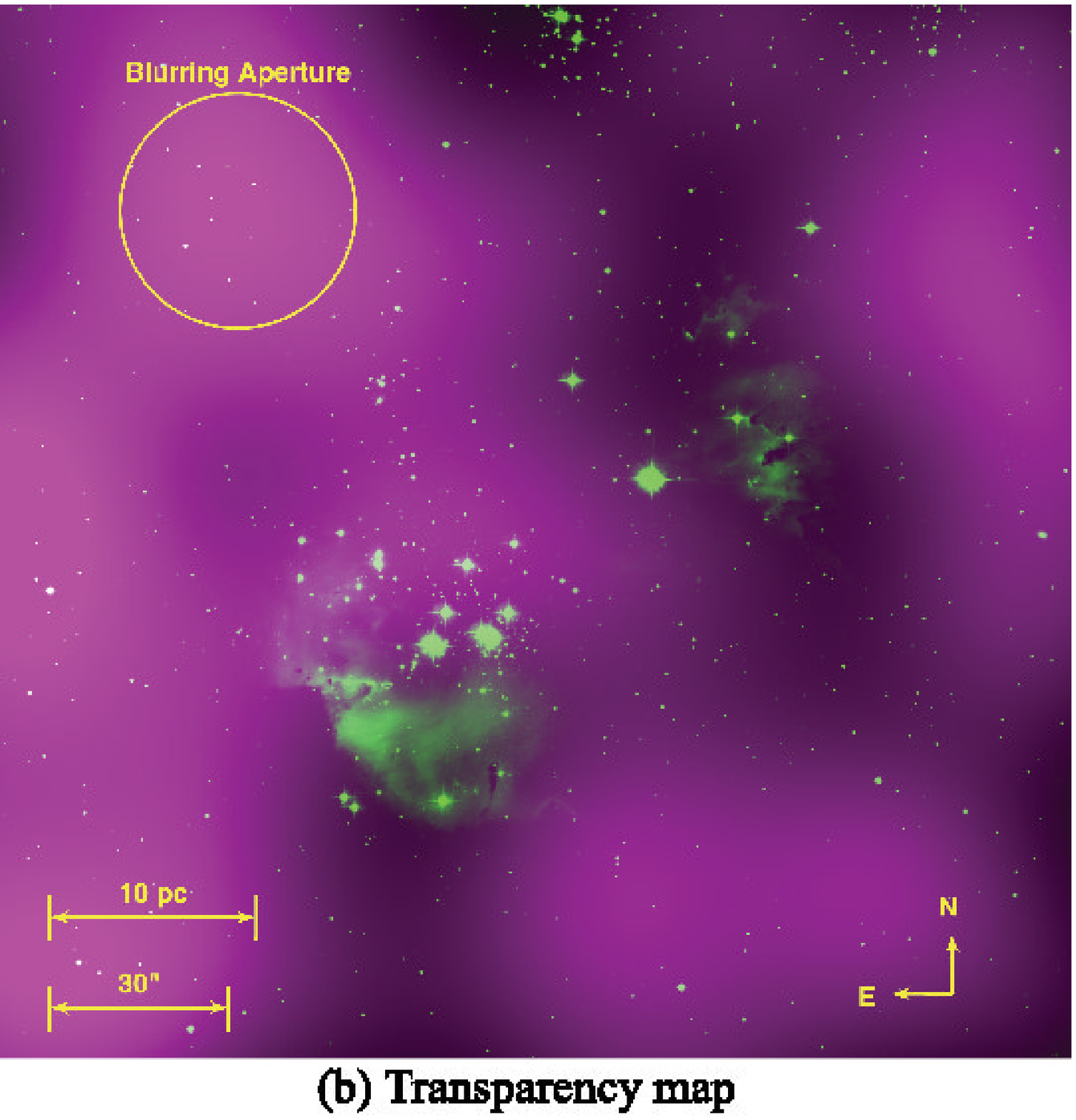}
\par\end{centering}

\caption{(a) Background objects visible in the NGC~602 region are identified
by circles overlaid on a V+I+H$\alpha$ composite image. The objects
were identified visually in the I-band image, required a corresponding
feature in V but none in H$\alpha$ and finally, confirmed as a non-stellar
object by SExtractor in I-band. The three sample objects identified
as A, B and C are expanded and shown in Figure~\ref{fig:Example-background-galaxies}.
(b) A transparency map created by mapping the density of detected
galaxies. The density map was obtained by convolving a map of detected
galaxy locations (each a single pixel with value of unity) with a
Gaussian kernel whose FWHM aperture is shown in the figure.
This image is shown in magenta (brighter is more transparent)
while the V-band ACS image is overlaid in green for reference. The
image intensity scale is arbitrary. \label{fig:Background-objects-visible} }

\end{figure}

\begin{figure}
\begin{centering}
\includegraphics[scale=0.7]{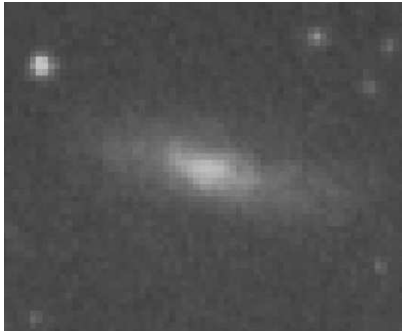} 
\par\end{centering}

\begin{centering}
\includegraphics[scale=0.7]{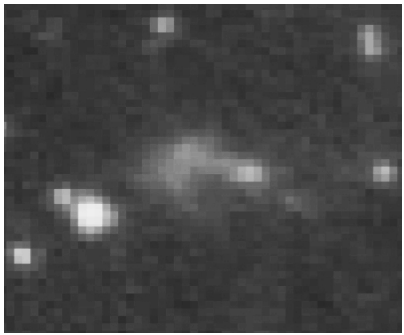} 
\par\end{centering}

\begin{centering}
\includegraphics[scale=0.7]{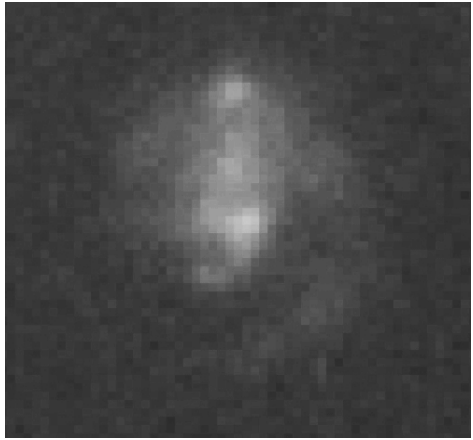} 
\par\end{centering}

\caption{Example background galaxies expanded in the central region shown in
I band only. Letters A, B and C correspond to locations identified
in Figure~\ref{fig:Background-objects-visible}.\label{fig:Example-background-galaxies}}

\end{figure}

The lack of large-scale dense gas and dust in the vicinity of the nebula
is established by a systematic search for background objects in the
ACS images. The results are shown in Figure~\ref{fig:Background-objects-visible}(a),
which shows the locations of detected galaxies throughout the nebula.
Background galaxies were located by first visually identifying extended
objects in the I-band image, then verifying a corresponding feature
in the V-band with no extended feature in the H$\alpha$ image. Finally, each object
was confirmed as non-stellar using SExtractor software \citep{bertin1996}.
Examples of three background galaxies found in the central region
are shown in Figure~\ref{fig:Example-background-galaxies} which
correspond to locations A, B and C in Figure~\ref{fig:Background-objects-visible}(a).
On scales on the order of the nebula, distant galaxies are relatively
evenly distributed across the image. The fact that there is a general
transparency throughout the nebula indicates that there is no large-scale,
dense cloud in the plane of the sky from which this nebula may have
formed. In addition, the transparency around the nebula beyond the
PDRs indicates that it did not break out from within a large-scale
high density molecular cloud complex, but more likely from a small, dense
region on the order of the nebula's size or smaller, $\sim$20~pc,
on the periphery of the southwest cloud.

On smaller scales, there are regions in Figure~\ref{fig:Background-objects-visible}(a)
that seem to have a distinct lack of detected galaxies, suggesting
an optically thick molecular gas substructure. An example is the region
to the south of the apparent molecular outcropping on the southeast
face. We would expect to find a clump or filament of molecular gas
in this region which would be eroded by the nearby group of OB stars.
This apparent substructure is revealed by plotting the density of
detected galaxies across the image with higher densities corresponding
to higher optical transparency.

First, an otherwise blank image was
formed with galaxy location pixels given a value of unity. This galaxy
detection map was then convolved with a Gaussian kernel producing a map
of relative galaxy density, i.e., transparency, shown in Figure~\ref{fig:Background-objects-visible}(b)
with the F555 (V-band) filter image superimposed to provide a reference
to the nebular features. The kernel used had a FWHM of $40\,\mbox{arcsec}$,
indicated by the circle in the image. The scale of the Gaussian kernel was chosen empirically,
with smaller scales too sensitive to individual galaxies and larger
scales suppressing the variation in density. This FWHM corresponds
to a physical scale of $12\,\mbox{pc}$ which, appropriately, is comparable
to the smallest scale of giant molecular clouds.

Although processing
an arbitrary field of galaxies in this manner would also produce an
apparent substructure, the dark regions marking obscured zones in
Figure~\ref{fig:Background-objects-visible}(b) lie very much where
we would expect them to. Most notably, there is a distinct column
extending southward from the apparent molecular outcropping on the
southeast face toward the region south of the image where there is
a high HI column density as shown in the lower image of Figure~\ref{fig:NGC-602-and-},
suggesting a physical connection via dense molecular filaments. There
is also evidence of a filament extending from the southwest corner
to the north central edge of the image, connecting to the HI region
to the south and tracing the northwest PDR and the cluster to the
north. We further establish this connection on a kinetic basis as
well in \S~\ref{sub:Gas-Kinematics}.

\subsection{Gas Kinematics\label{sub:Gas-Kinematics}}

The maximum HI column density within a 250~pc radius of the nebula is
3.4$\times$10$^{21}\,$~cm$^{-2}$. For the SMC HI survey, the column
density in the optically thin regime was found to be $N_{HI}$~$<$~2.5$\times$10$^{21}\,$cm$^{-2}$
\citep{stanimirovic1999}. Since this is only slightly higher than the optically
thin limit, we can assume that in this region, we are sampling the
greater portion of the neutral hydrogen in the line of sight. There
is clearly a relative HI void to the north of the nebula and relatively
dense HI to the south. The position of NGC~602 on the edge of the
dense HI region suggests that its progenitor molecular cloud may have formed out of the HI cloud
structure to the south.

\begin{figure}
\begin{centering}
\includegraphics[width=8cm]{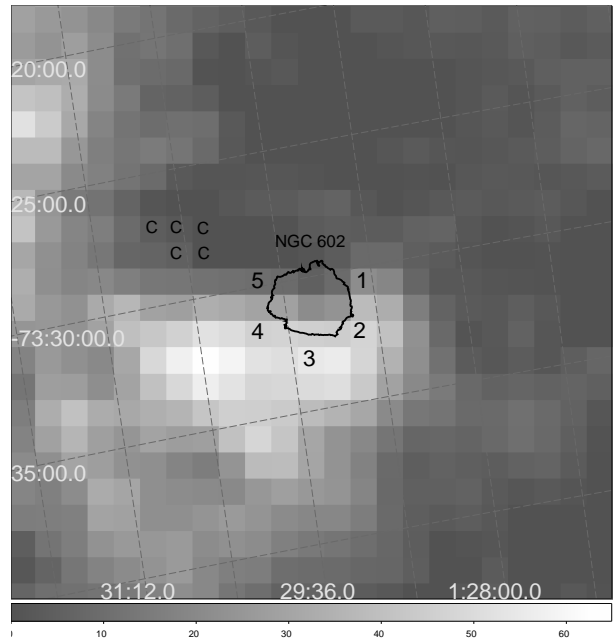} 
\par\end{centering}

\caption{The data cube slice at $179.2\,\mbox{km\, sec}^{-1}$ heliocentric
velocity where HI intensity peaks south of the NGC~602 HII region
(outlined) and correlates with the HII velocity across the nebula.
The numbered pixels correspond to velocity profile samples around
the nebula. The pixels labeled with ``C'' have been identified as
having a profile representative of a baseline common to all pixels
in the area. Coordinates are J2000. The intensity scale is
brightness temperature in K within
the $1.6\,\mbox{km\, sec}^{-1}$ channel.\label{fig:Cloud-Velocity}}

\end{figure}

\begin{figure}
\begin{centering}
\includegraphics[width=80mm]{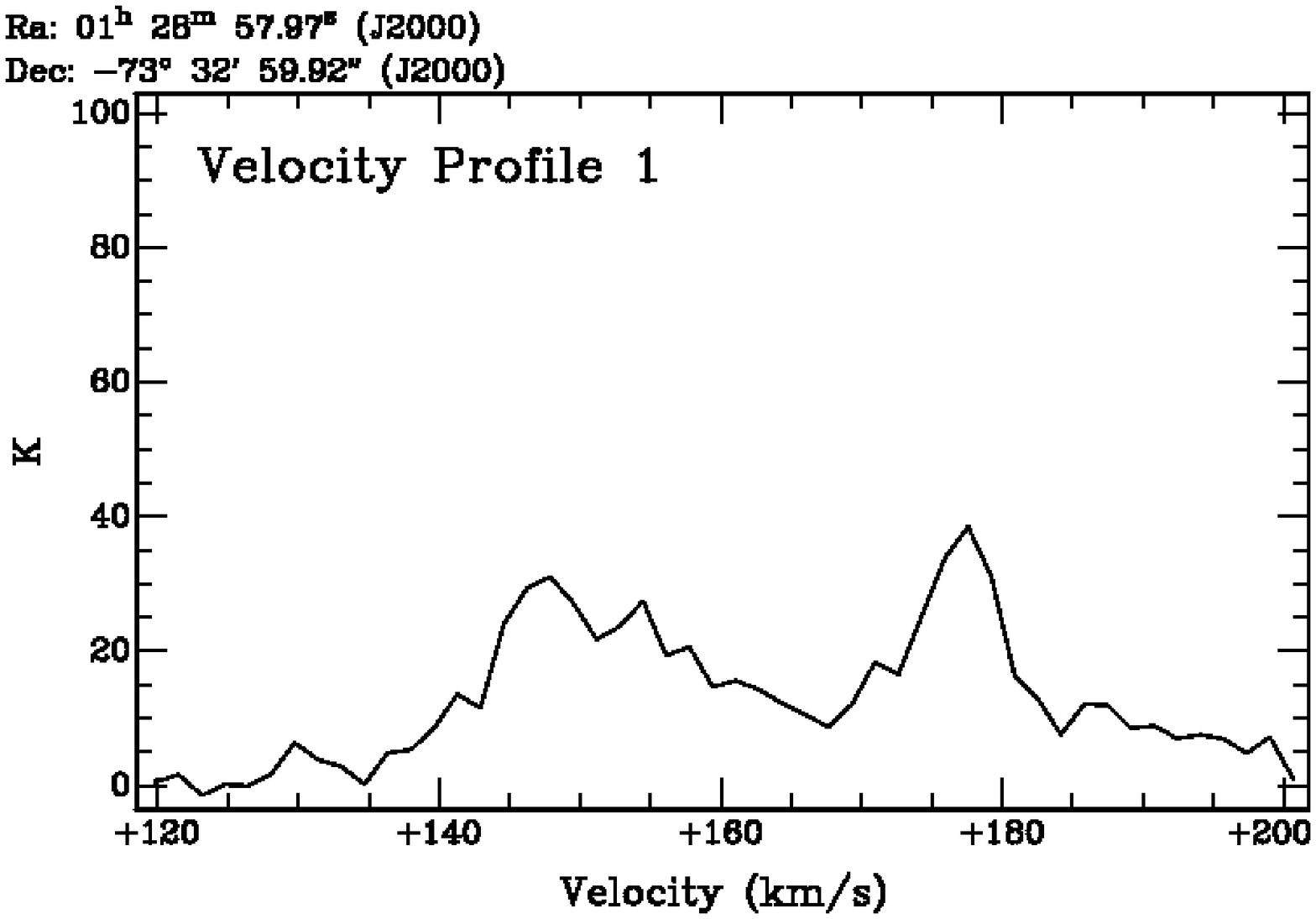} 
\par\end{centering}

\begin{centering}
\includegraphics[clip,width=80mm]{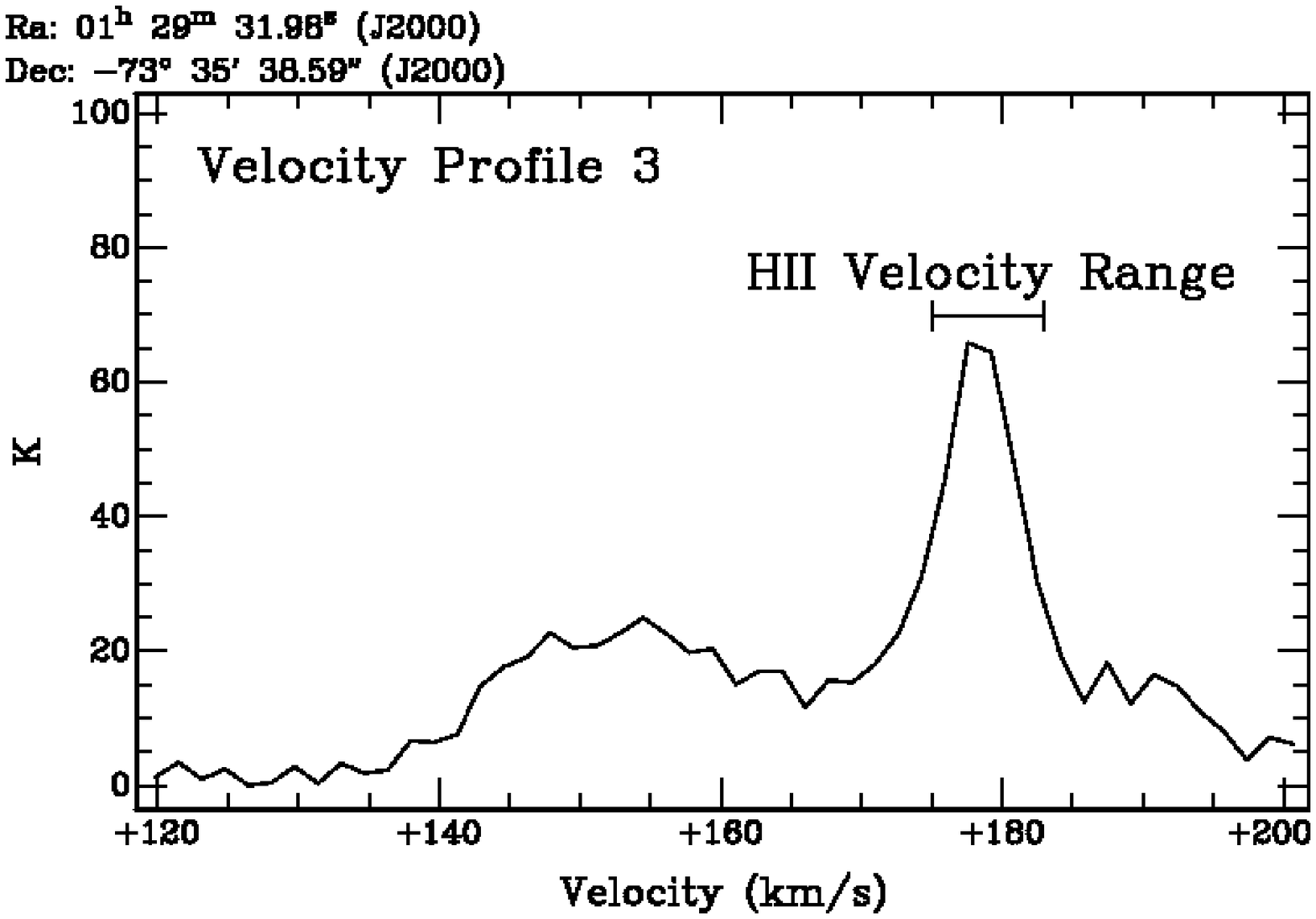} 
\par\end{centering}

\begin{centering}
\includegraphics[width=80mm]{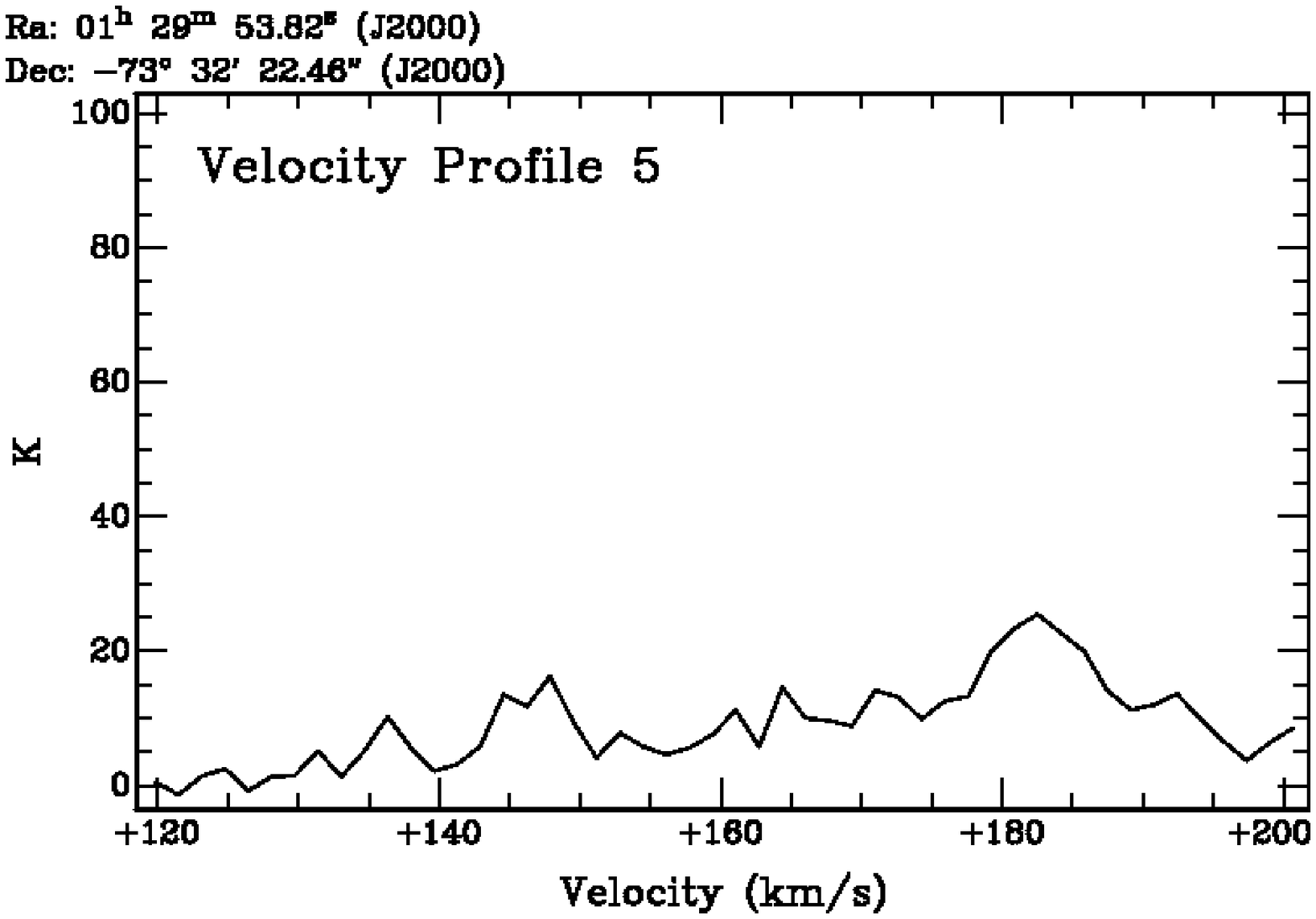} 
\par\end{centering}

\caption{HI velocity profiles sampled around NGC~602. Location numbers correspond
to those in Figure~\ref{fig:Cloud-Velocity}. The HII velocity dispersion
found across the nebula is indicated by the bar above the component
in profile 3.\label{fig:HI-velocity-profile}}

\end{figure}

To look for a correspondence, we explored the HI data cube visually
by stepping through velocity slices and looking for emission peaks
near the nebula. We found a peak of HI to the south-southwest which
moved along the lower periphery of the nebula as the central velocity
of the slice increased, and diminished when it reached the south-southeast.
The effect is that of a clump of HI with a narrow velocity range having
a morphology such that it \char`\"{}hugs\char`\"{} the southern half
of the nebula. The fact that the velocity sequence traced the lower
boundary of the nebula implies a specific connection between the nebula
and the cloud. The middle of this HI velocity sequence is
179~km~sec$^{-1}$ (heliocentric) which is shown
in Figure~\ref{fig:Cloud-Velocity}. Numbered pixels correspond to
velocity profiles sampled around the nebula, some of which are shown
in Figure~\ref{fig:HI-velocity-profile}. This sequence of plots
shows relatively weak components to the west (HI Profile 1) and east
(HI Profile 5), with a strong narrow component at the southern periphery
(HI Profile 3). The peak velocity increases from west to east as we
observed visually. The narrow component in HI Profile 3 has a FWHM
of about 6~km~sec$^{-1}$.

\subsection{Progenitor Cloud}

Having identified a component of the HI spectrum as a candidate source
of progenitor gas for NGC~602, we isolated it by first subtracting a baseline
spectral component common to all the pixels in the region. The baseline
component spectrum was identified by noting that a group of 5 pixels
to the northeast of NGC~602 have a broad baseline spectrum without
any discrete components. This baseline spectrum also appears to be
a common component to the spectra at all pixels in the vicinity. These
pixels are labeled \char`\"{}C\char`\"{} in Figure~\ref{fig:Cloud-Velocity}.
These baseline HI spectra were averaged and globally subtracted as
continuum from the NGC~602 region HI spectrum. The desired cloud
component was then defined by making a velocity cut from 170~to~185~$\mbox{km}\,\mbox{sec}^{-1}$.

\begin{figure}
\begin{centering}
\includegraphics[width=8cm]{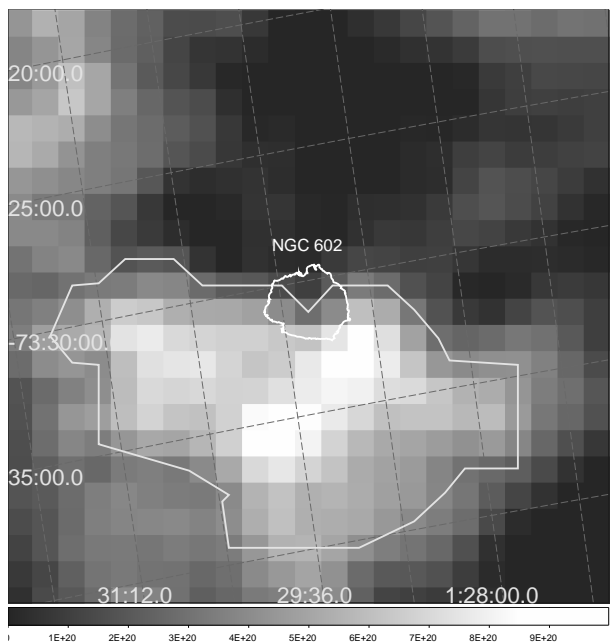} 
\par\end{centering}

\caption{Boundary of isolated cloud component HI~J0130-7337+180H
used in progenitor cloud calculations.
The image represents the integrated column density after removing
the baseline continuum component and making a velocity cut from 170
to 185 $\mbox{km\, sec}^{-1}$. Coordinates are J2000. The intensity
scale is in units of cm$^{-2}$.\label{fig:Isolated-cloud-component}}

\end{figure}

The integrated column density derived from the resulting cloud component
is shown in Figure~\ref{fig:Isolated-cloud-component}. As expected,
there is a distinct clump of enhanced column density south of the
nebula which is outlined in the figure. This is NGC~602's progenitor
cloud, hereafter referred to as {HI~J0130-7337+180H}%
\footnote{The designation HI~J0130-7337+180H describes an HI cloud component
at J2000 coordinates 01h30m~,~$-73^{\circ}$37' with +180 km~sec$^{-1}$~heliocentric
radial velocity.%
}. Partially visible is a filament extending from the northeastern
corner of this cloud well away from NGC~602. The HI column density
within the boundary shown was integrated and multiplied by an assumed
mean atomic mass, $\overline{m}=1.5m_{H}$, and the cloud was then
modeled as a sphere of equivalent mass. There is insufficient data
to estimate the spin temperature of the cloud, but \citet{Dickey2000}
combined emission and absorption spectra at background source points
across the SMC to obtain a sample of cold cloud phase temperatures
with a range of about 10~to~150~K. The temperature of {HI~J0130-7337+180H}
is assumed to be in this range. The resulting parameters for the idealized
cloud are shown in Table \ref{tab:CloudParameters}.

\subsection{HI Shell Structure}\label{sub:HI-Shell}

Referring to shell structure shown in Figure~\ref{fig:HI-total-column},
NGC~602 is located inward of the projected intersection of HI shells
497 and 499, from \citet{staveley-smith1997}, which are both expanding
toward the south-southwest and NGC~602 is situated just inward of
the overlap area. Depending on their positions along the line of sight,
these two shells could have interacted in the overlap region as they
expanded. Given that their centers are near each other on the sky,
if they are also in proximity radially, this interaction could have
been continuous over a good portion of their history. A super-giant HI
shell, SGS~494A from \citet{stanimirovic1999}, also potentially could interact in recent times with
NGC~602 as it expands in the opposite direction from the south-southwest.
The parameters of the HI shells in the NGC~602 region are given in
Table \ref{tab:Characteristics-of-HI}. NGC~602 is in the region
where at least three HI shells may interact.

\subsection{HII Region Velocities}

\begin{figure}
\begin{centering}
\includegraphics[width=8cm]{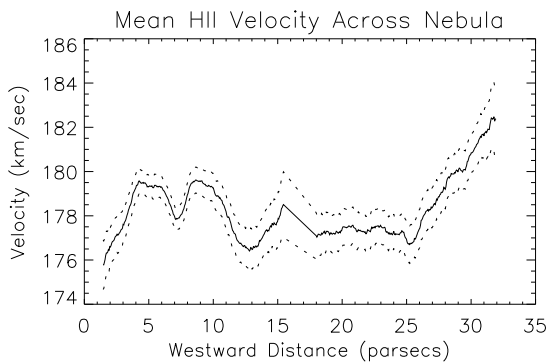} 
\par\end{centering}

\caption{The mean HII velocity derived from the two slits, (solid line) smoothed
to $\sim$1~pc with approximate position along the nebula
from east to west as the horizontal axis. Distorted data due to edge
effects on the spectra and from smoothing is suppressed. Linear interpolation
was used to fill the resulting gap in the middle of the plot. The
dashed lines represent the centroiding error limits (1$\sigma$).
Including error, the minimum and maximum velocities are 174.7 and
184.1~km~sec$^{-1}$, respectively.\label{fig:Slits}}

\end{figure}

In order to understand the nebula's relationship to the surrounding
HI spatial and velocity structure, we derived radial velocities of
the ionized gas across the nebula from the AAT UCLES H$\alpha$ spectroscopy
described in \S~\ref{sec:Data}. The mean H$\alpha$ emission line wavelength
was determined by centroiding at each position along the two slits,
shown in Figure \ref{fig:NGC602-Halpha-Slits}, which together span
the width of the nebula. Figure \ref{fig:Slits} is a plot of the
mean heliocentric H$\alpha$ velocity across the nebula derived from these
spectra and horizontally scaled in parsecs.
The radial velocity covers a range from 174.6~to~184.2~km~sec$^{-1}$,
and the RMS dispersion in radial velocity of the line peak is only
1.4~km~sec$^{-1}$. The nebula appears to be quite quiescent.

The H$\alpha$ velocity range across the nebula from the longslit
spectra is indicated with a bar above the middle plot in Figure~\ref{fig:HI-velocity-profile}.
Note that the range is nearly centered with the peak of the narrow
southerly HI cloud component and is well-constrained within its range.
This strong correspondence between the radial velocity of the ionized
gas and position of a prominent HI cloud component in the complex
south of the nebula shows that NGC~602 is co-moving with this HI
cloud component and therefore likely formed from it.

\section{Discussion}

The part of the Wing of the SMC where NGC~602 resides is characterized
by a moderate HI column density of 3~to~4$\times10^{21}$~cm$^{-2}$
and a relatively simple HI shell structure. This total column density
is approximately 3~to~4 times the HI star formation threshold for
faint dwarf galaxies found by \citet{begum2006}, so in general, star
formation might be expected, but is not strongly indicated. The formation
of one group of OB stars such as NGC~602 in this environment is therefore
not surprising, but the simplicity and diffuse nature of its surrounding
ISM structure, its relative isolation and sparse stellar population
lends itself to a straightforward evaluation and comparison of formation
scenarios.

\subsection{Environment}

The low degree of optical obscuration indicated by the presence of galaxies throughout
the N90 nebula and its surroundings is consistent with the low levels
of extinction in the SMC Wing reported by \citet{Gordon2003}. In
that study, the HI column density-to-extinction ratio was sampled
at AzV~456, an OB star $1.6^{\circ}$ to the northwest of NGC~602,
and was found to be\begin{equation}
N(HI)\bigr/A(V)=(7.4\pm1.20)\times10^{21}\ \mbox{cm}{}^{-2}\,\mbox{mag}^{-1}.\end{equation}

The HI column density in the region of the HST image ranges from 1.3$\times$10$^{21}\,\mbox{cm}^{-2}$~
in the northernmost central point to $3.1\times10^{21}\,\mbox{cm}^{-2}$
in the far southwest corner. Using this relationship, the corresponding
expected extinction in the visual band (i.e., in the F555W band) ranges
from 0.7~to~1.7~mag. This is not enough to significantly suppress
detection of galaxies, as evident in the HST images. However, as illustrated
by Figure \ref{fig:Background-objects-visible}(b), a locally dense
substructure not visible in the lower resolution HI map connects the
nebular structures to {HI~J0130-7337+180H}, which
is likely a remnant of the progenitor molecular cloud.

\subsection{Kinematics}


Although clearly interacting with its environment, NGC~602 has apparently not
produced a significant amount of mechanical energy since it has not
significantly disrupted any higher density large-scale HI structures.
This is consistent with the young age of the stars which may not have
produced any supernovae (see \citealp{smith2008}), and the expected
low mechanical energy in stellar winds from the OB stars of NGC~602
given the low metallicity of the SMC \citep{walborn2000}. This lower
metallicity leads to reduced stellar wind mass loss rates, which
in some cases can be dramatic (e.g., \citealp{bouret2003}).

The gentle velocity profile of the ionized gas in Figure~\ref{fig:Slits}
supports this picture of relatively peaceful birth and evolution of
the gas associated with NGC~602. This cluster is
ionizing the diffuse inter-cloud gas and illuminating the surfaces
of large-scale structures, while creating smaller-scale structure
along the PDRs on cloud peripheries through photoionization, leading
to the rich fine structure seen in the HST images (e.g., Figure \ref{fig:NGC602-Halpha-Slits}).

\subsection{Origins}

We have established that the morphology and kinematics of the N90 nebula is 
consistent with cluster formation 
through a blistering process in the periphery
of {HI~J0130-7337+180H}. In order to verify that NGC~602's mass
is consistent with the cloud's mass, we estimate an upper limit to the 
star formation efficiency. The total stellar
mass has been estimated by \citet{cignoni2008} at $\sim2000M_{\odot}$ , corresponding to
roughly $\sim0.7\%$ star formation efficiency based on the cloud's HI mass 
estimate given in Table \ref{tab:CloudParameters}. This is only an upper limit
because we do not know how much molecular gas exists in the region, but the limit is
reasonable.
Given that the cluster formed from the cloud, we now consider what could have led to the 
conditions that allowed core formation and collapse within it.

For the formation of a gravitationally bound molecular region through collapse within the HI cloud, the Jeans length 
must be smaller than the HI cloud's scale and the the cloud must be undisturbed for a 
period on order of
the free-fall time, $t_{ff}$. Assuming {HI~J0130-7337+180H}'s precursor was uniform
and its global characteristics were 
similar to those of Table \ref{tab:CloudParameters}
prior to collapse,
the Jeans length was less than 15~pc. This is well within
the  220~pc scale of the cloud, indicating that bound regions could have formed through simple
gravitational collapse. However the time required is
$t_{ff}\sim$40~Myr, during which time the cloud would have to
be undisturbed by significant interactions. This is not likely 
considering the dynamics of clouds in the SMC. The radial velocity dispersion of shells
in the Wing region is 19.4~km~sec$^{-1}$
(see Table \ref{tab:Characteristics-of-HI}), or 27~km~sec$^{-1}$ transversely on the sky.
Although a shell structure is enforced to obtain this value, this
would also be the approximate cloud dispersion velocity
regardless of the model (i.e., pure turbulence). At this speed, the
cloud would need to travel 1.1~kpc or $\sim$1$^\circ$ on the sky
(about half the scale of the
SMC Wing) without a significant
interaction to experience Jeans instability molecular cloud formation. Since there is insufficient time
for spontaneous collapse, some type of interaction is required
to accelerate the collapse.

A possible mechanism to stimulate collapse is a gas flow.
The HI shell fragments in the region may have been exposed
to continuous interactions with their surroundings, encouraging the formation of turbulence.
Such a process is described in \citet{slavin1993} where hot gas flowing
past a cool gas cloud can produce an intermediate warm turbulent mixing
layer. If, under some conditions, a similar process could lead to further cooling and
subsequent star formation in the cloud, it would do so on its periphery.
The morphology of the region with NGC~602 forming on the boundary between
a cool HI cloud and the large diffuse region to the north seems consistent with
this scenario. Interestingly, \citet{Hoopes2002} detected a hot gas component
in this sector of the SMC through OVI absorption studies. The velocity spectrum of the
hot gas can be interpreted as consistent with a hot outflow or possibly as
hot gas flowing through the diffuse HI gas of the Wing.

\begin{figure}
\begin{centering}
\includegraphics[width=80mm]{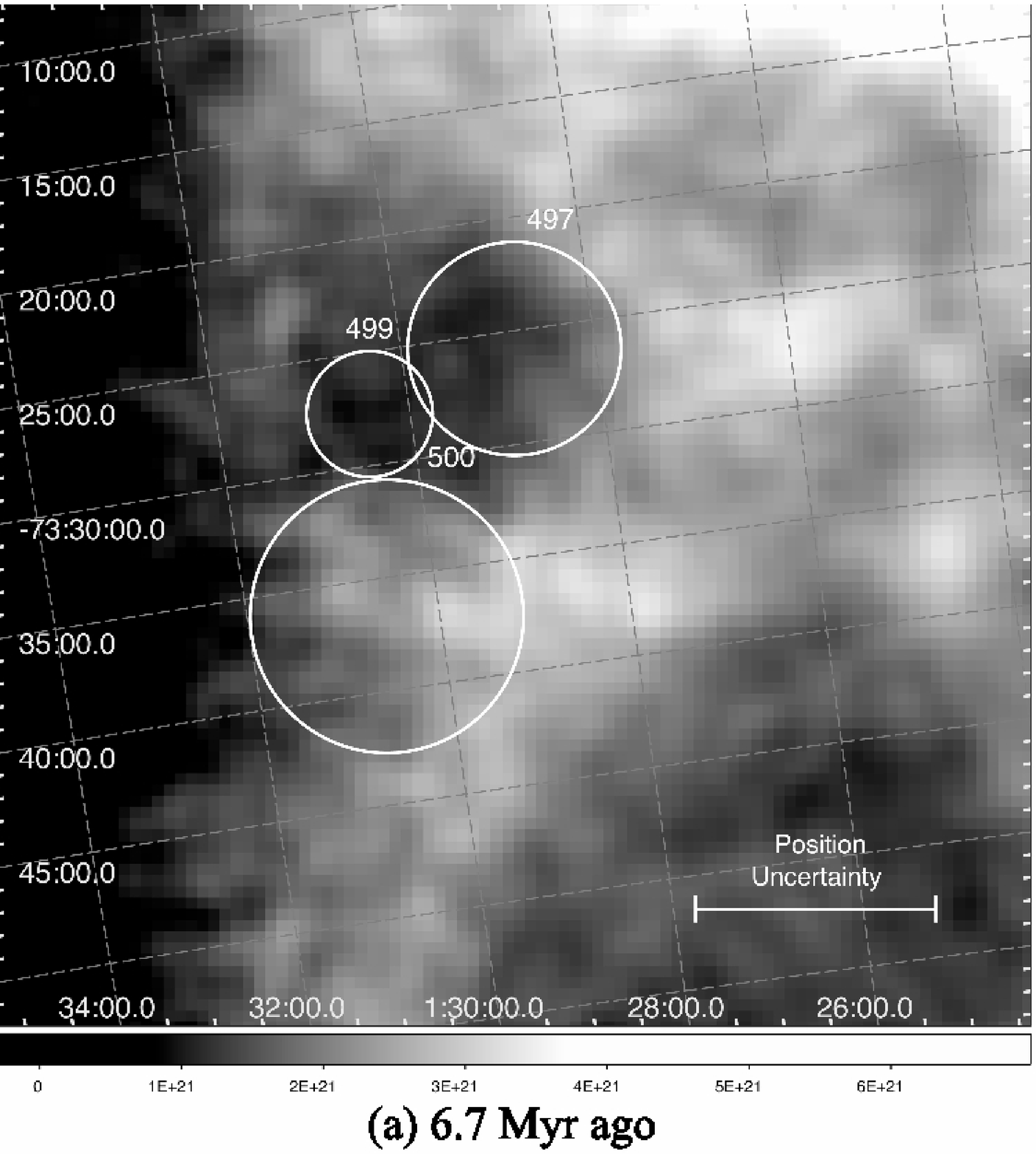}
\includegraphics[width=80mm]{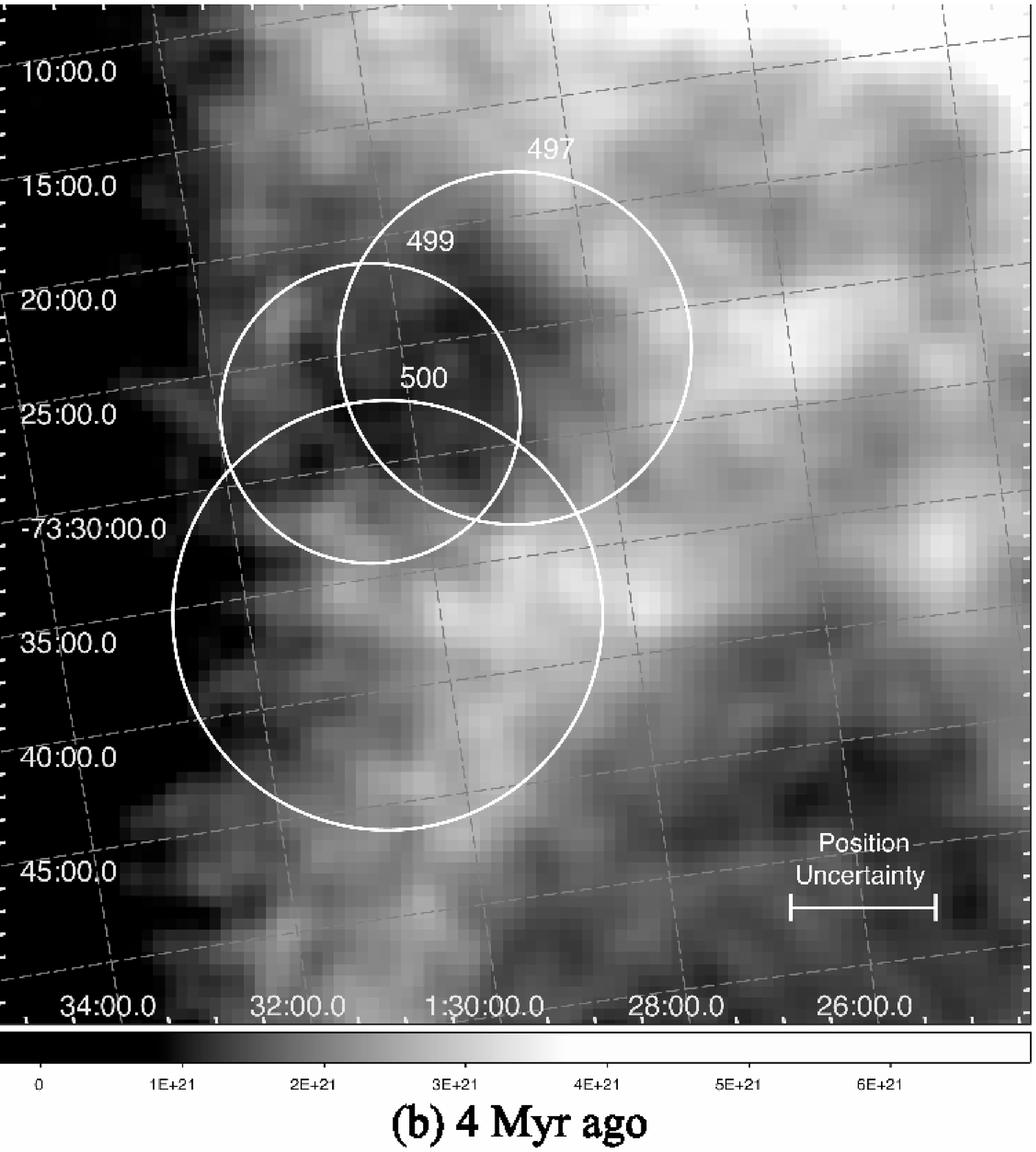}
\par\end{centering}

\begin{centering}
\includegraphics[width=80mm]{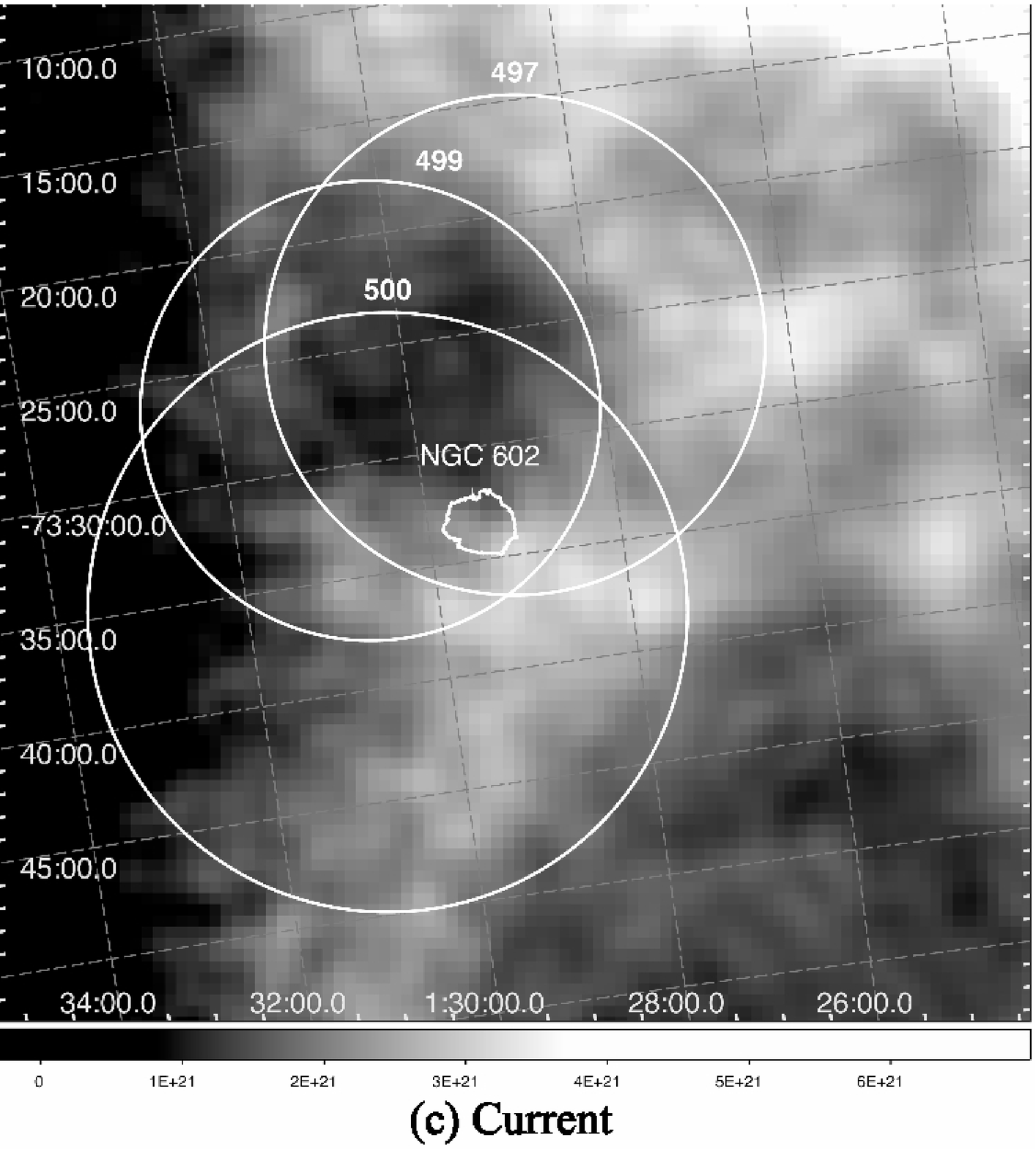} 
\par\end{centering}

\caption{Estimated shell evolution and possible interaction leading to NGC~602
formation. The estimated transverse shell position uncertainty is
indicated by the bar in each diagram. This is calculated as $\sigma_{t}=\sqrt{2}\sigma_{vr}t$
where $\sigma_{vr}=$19.4~km~sec$^{-1}$ is the radial velocity
dispersion in the SMC Wing region \citep{staveley-smith1997}. Coordinates
are J2000. The intensity scale is in units of cm$^{-2}$.\label{fig:Estimated-shell-evolution}}

\end{figure}

A perhaps more straightforward 
mechanism for the required gas interactions are the expanding HI shells discussed 
in \S~\ref{sub:HI-Shell}, which would collide and overlap as they expand.
The radii of shells in a snowplow phase are approximately 
$R(t)=R\Bigl(t_{0}\Bigr)\left(\frac{t}{t_{0}}\right)^{0.55}$, where
the exponent is the average of that for energy- and momentum-conserving modes.
The average is used
here to simplify approximating the radii. $t_{0}$ is the current
dynamical age of the shell as shown in Table \ref{tab:Characteristics-of-HI}.
In Figure~\ref{fig:Estimated-shell-evolution} the three shells are 
shown at the time of their first interaction,
at the estimated formation of NGC~602 and at the present time. The earliest interaction
occurs at $\sim$6.7~Myr between shells 497 and 499, assuming their
centers are radially at the same position. As they expand, the overlapping
region of the shells propagates to the southwest directly toward the
current position of the nebula. At 4~Myr, when the NGC~602 cluster
began to form, the shells would have been interacting for $\sim$2.7~Myr 
and may have also had a brief interaction with shell 500. Turbulence, 
compression and radiative cooling produced in this interaction may have 
been sufficient to lead to rapid star formation.

A quantitative test of this model would
require hydrodynamic simulations of the shell interactions, but potentially 
relevant 3-dimensional numerical simulations have been
performed by \citet{heitsch2008}. Their study showed that uniform,
low density warm gas flows in direct collision can lead to instabilities
resulting in turbulence and rapid cooling, followed by cloud formation. The initial conditions
considered in that study were $n=3$~cm$^{-3}$, $T=1800$~K
and $v_{inflow}=7.9$~km~sec$^{-1}$. Interaction
times on the order of 3~Myr were sufficient to cool the warm gas flows in the
collision region to a point where $\sim$50\% of
the gas was below 100~K and for 
dense substructures to form, which plausibly could support star formation.

The initial temperature, density and velocity of the flows in these simulations
are on the same order as might be found in the much younger shells at the time they began to
interact $\sim$7~Myr ago. The 3~Myr cooling time found in the simulations
is comparable to the estimated 2.7~Myr allowed to establish
conditions for star
formation in this shell interaction scenario. Although the interaction is not a head-on collision of flows as in
the simulations, there are similarities. When two shells begin to interact,
the interaction
region forms an expanding ring (the intersection of the two spherical shells)
which continues to expand until the shells dissipate. The collision
region appears locally at any point on the ring as two slabs
crossing obliquely and continuously. The obliquely crossing gas ring,
therefore, has a head-on
velocity component that emulates the simulated interacting flows and also a
parallel component which carries the interacting gas with it, expanding the
ring.

\section{Conclusions}

The young star cluster NGC~602 and its associated HII region N90
formed in relative isolation in the Wing of the SMC under conditions
that are generally marginal for star formation, the ISM in this region
being relatively diffuse and quiescent. A general transparency to
galaxies throughout the nebula and its surroundings shows that there
are no large-scale dust and molecular concentrations in or near the
nebula to explain its formation through simple gravitational collapse.
There is also no morphological evidence of violent events such as
supernovae or strong stellar winds having directly triggered its initial
formation as demonstrated by the low radial gas velocity gradient
and dispersion measured across the nebula. Our measured low velocity
gradient across the nebula also shows that NGC~602's formation and
evolution has also been peaceful and, to our good fortune, has left
the evidence of its birth relatively undisturbed. Star formation triggered
by gas flow interaction might provide a consistent explanation for
NGC~602's formation under these constraints.

Based on a tight, unambiguous neutral and ionized gas velocity correlation
at $\sim$180 $\mbox{km\, sec}^{-1}$, we determined that NGC~602
formed on the periphery of a low-density HI cloud component
immediately to its south and blistered from it. The HI density of
the natal cloud, designated {HI~J0130-7337+180H} was estimated at
1.3~cm$^{-3}$ and its size at 220~pc implying a mass of $3\times10^{5}M_{\odot}$.
This mass is sufficient to provide NGC~602's approximate stellar mass with
$\leq0.7\%$ star formation efficiency. Given the temperature range of cool
cloud components in the SMC of 10~K~to~150~K,
the cloud's size is comparable to the Jeans length for a simple gravitational
collapse scenario. The timescale for undisturbed collapse
is $\sim$40~Myr, much longer than the dynamics of the associated
HI shells in the SMC would allow. The interaction of large-scale gas
components likely accelerated the formation process, producing
local molecular concentrations from which the cluster formed. The
pattern of transparency to galaxies in the region indicate the presence
of fragments of these molecular components in and around the nebula
connecting nebular features to the natal cloud.

A simple model of local HI shell evolution shows that two of the expanding
shells would have begun interacting $\sim$7~Myr ago
and that the nominal path of this interaction region is consistent
with the location of NGC~602. Turbulence at the intersection could
have led to the formation of this cluster $\sim$3~Myr later. Further
study of the NGC~602 region thus offers an opportunity to explore
star formation processes in a low density, relatively quiescent environment.

\acknowledgements{}

Support for program G0-10396 was provided by NASA through a grant
from the Space Telescope Science Institute, which is operated by the
Association of Universities for Research in Astronomy, Inc., under
NASA contract NAS 5-26555.

We thank Lynn Carlson for her work in determining the age of the NGC~602
cluster. We thank Fabian Heitsch for his insights into the applicability of colliding
flow simulations. We are grateful to the anonymous referee for helping us
improve this paper through very useful comments. JSG, SS and LN thank
the University of Wisconsin Graduate School for their partial support.

\clearpage

\begin{center}
\begin{deluxetable}{ccc}
\tablewidth{0pt}
\tablecaption{Parameters for SMC cloud HI~J0130-7337+180H\tablenotemark{a}\label{tab:CloudParameters}}
\startdata
\tabularnewline
\hline
\hline
Diameter&d&$220$~pc\\
HI density&$n_{HI}$&$1.3$~cm$^{-3}$\\
Mass density&$\overline{\rho}$&$3.2 \times 10^{-24}$~gm$\times$cm$^{-4}$\\
Mass&M&$2.7 \times 10^5$~M$_{\odot}$\\
Spin Temperature&T&$10$--$150$~K\\
\enddata
\tablenotetext{a}{Modeled as a uniform sphere.}
\end{deluxetable}
\par\end{center}

\begin{center}
\begin{deluxetable}{ccccccc}
\tablewidth{0pt}
\tablecaption{Characteristics of HI shells near NGC 602\tablenotemark{a}\label{tab:Characteristics-of-HI}}
\rotate
\startdata
\tabularnewline
\hline
\hline
& Center Distance & Center Angle & & Heliocentric & Expansion & Dynamical\\

Shell ID &
from NGC 602\tablenotemark{b} &
wrt NGC 602\tablenotemark{b} &
Radius\tablenotemark{c} &
Velocity &
Velocity &
Age\\

& (pc) & ($^\circ$E~of~N) & (pc) & (km$\times$sec$^{-1}$) & (km$\times$sec$^{-1}$) & (Myr)\\
\hline
SGS~494A& 670& 193.3& 650/540& 159& 28& 12\\
497& 150& 341.0& 194& 180.3& 13.4& 8.5\\
499& 130& 32.3& 179& 160.7& 14.2& 7.4\\
500& 90& 121.6& 233& 163.0& 15.4& 8.8\\
\enddata
\tablenotetext{a}{From \citet{staveley-smith1997} and \citet{stanimirovic1999}.}
\tablenotetext{b}{NGC 602 position is RA~1:29:30,~DEC~$-$73:34:00~(J2000).}
\tablenotetext{c}{Radius for spherical models, semi-major/semi-minor axes for ellipsoidal models.}
\end{deluxetable}
\par\end{center}

\clearpage

\begin{thebibliography}{}
\bibitem[Abel et al.(2007)]{abel2007} Abel, T., Wise, J.~H., Bryan,
G.~L.~2007, \apjl, 659L, 87A

\bibitem[Begum et al.(2006)]{begum2006} Begum, A., Chengalur, J.~N.,
Karachentsev, I.~D., Kaisin, S.~S., \& Sharina, M.~E.\ 2006, \mnras,
365, 1220

\bibitem[Bertin and Arnouts(1996)]{bertin1996} Bertin, E., \& Arnouts, S.\ 1996, \aaps, 117, 393

\bibitem[Bouret et al.(2003)]{bouret2003}Bouret, J.~C., Lanz, T.,
Hillier, D.~J., Heap, S.~R., Hubeny, I., Lennon, D.~J., Smith,
L.~J., Evans, C.~J., 2003, \apj, 595, 1182

\bibitem[Carlson et al.(2007)]{carlson2007}Carlson, L.~R., Sabbi,
E., Sirianni, M., Hora, Nota, A., Meixner, M., Gallager, J.~S.,~III,
Oey, M.~S., Pasquali, A., Smith, L.~J., Tosi, M., Walterbos, R.\ 2007,
\apj, 665, L109

\bibitem[Cignoni et al.(2008)]{cignoni2008}Cignoni, M. et al.~2008, in preparation.

\bibitem[Dickey et al.(2000)]{Dickey2000}Dickey, J.~M., Mebold,
U., Stanimirovic, S., \& Staveley-Smith, L.\ 2000, \apj, 536, 756

\bibitem[Gordon et al.(2003)]{Gordon2003}Gordon, K.~D., Clayton,
G.~C., Misselt, K.~A., Landolt, A.~U., \& Wolff, M.~J.\ 2003,
\apj, 594, 279

\bibitem[Gouliermis et al.(2007)]{gouliermis2007}Gouliermis, D.~A.,
Quanz, S.~P., Henning, T.~2007, \apj, 665, 306

\bibitem[Heitsch et al.(2008)]{heitsch2008}Heitsch, F., Hartmann,
L.~W., Slyz, A.~D., Devriendt, J.~E.~G., \& Burkert, A.\ 2008,
\apj, 674, 316

\bibitem[Hilditch et al.(2005)]{hilditch2005}Hilditch, R.~W. , Howarth,
I.~D. \& Harries, T.~J.~2005, \mnras, 357, 304

\bibitem[Hodge(1983)]{hodge1983}Hodge, P.~W.~1983, \apj, 264,
470

\bibitem[Hoopes et al.(2002)]{Hoopes2002}Hoopes, C.~G., Sembach,
K.~R., Howk, J.~C., Savage, B.~D., \& Fullerton, A.~W.\ 2002,
\apj, 569, 233

\bibitem[Hutchings et al.(1991)]{hutchings1991}Hutchings, J.~B.,
Cartledge, S., Pazder, J., \& Thompson, I. B.~1991, \aj, 101, 933

\bibitem[Meaburn(1980)]{meaburn1980}Meaburn, J.~1980, \mnras, 192,
365

\bibitem[Slavin et al.(1993)]{slavin1993}Slavin, J.~D., Shull, J.~M.
\& Begelman, M.~C.,~1993, \apj, 407, 83

\bibitem[Schmalzl et al.(2008)]{schmalzl2008}Schmalzl, M., Gouliermis,
D.~ A., Dolphin, A.~E., Henning, T.,~2008, ArXiv e-prints, 0804,
0543

\bibitem[Smith et al.(2008)]{smith2008}Smith, L. J.~2008, in preparation.

\bibitem[Stanimirovi{\'c} et al.(1999)]{stanimirovic1999}Stanimirovi{\'{c}},
S., Staveley-Smith, L., Dickey, J.~M., Sault, R.~J., \& Snowden,
S.~L.\ 1999, \mnras, 302, 417

\bibitem[Staveley-Smith et al.(1997)]{staveley-smith1997}Staveley-Smith,
L., Sault, R.~J., Hatzidimitriou, D., Kesteven, M.~J., \& McConnell,
D.\ 1997, \mnras, 289, 225

\bibitem[Walborn et al.(2000)]{walborn2000}Walborn, N.~R., Lennon,  D.~J., Heap, S.~R., Lindler, D.~J., Smith, L.~J., Evans, C.~J.,  \& Parker, J.~W.\ 2000, \pasp, 112, 1243 

\bibitem[Westerlund(1964)]{westerlund1964}Westerlund, B.~1964, \mnras,
127, 149

\end{thebibliography}
\end{document}